# A continuous plane of polarization rotator and detector based on the liquid crystal Θ-cell

Deepak Kararwal, Rahul Panchal and Aloka Sinha*

*Department of Physics, Indian Institute of Technology Delhi, New Delhi, India.*
*\*aloka@physics.iitd.ac.in*

**Abstract:** Modulation of the polarization state of optical beams is crucial in advanced optical applications such as polarization microscopy, polarization rotation, and medical imaging. We demonstrate a continuous polarization rotator by utilizing the liquid crystals (LC) in Θ-cell configuration comprising of circular and parallel alignment of LC on two aligning glass substrates. A rotation of plane of polarization up to ~160˚ is observed using Stokes polarimetry and substantiated by the mathematical analysis using Mueller matrix formalism. Further, the device is demonstrated as a plane of polarization detector with high precision detection angle. The plane of polarization is analyzed through single shot imaging of the intensity distribution of transmitted light through the LC Θ-cell.

## 1. Introduction

The necessit*y* for rotation and detection of the plane of polarization is crucial in modern technology owing to the numerous practical applications in the domains of microscopy, medical imaging, beam steerers, and edge detection in optical imaging [1–7]. The phase or amplitude difference between the two orthogonal polarization components modulates the state of polarization (SoP) of light. Many devices utilize the phase retardation to control the PoP of the light [8–12]. Traditionally, half-wave retarders and Faraday rotators are used to control the PoP of light [13,14]. The conventional birefringent crystal-based retarders are mechanically controlled and expensive, while the Faraday rotators are limited by their large size, non-reciprocity, and rotation angles [14]. The liquid crystal (LC) media is also utilized for polarization rotation because of their high birefringence and dielectric anisotropy [15–17]. Generally, the rotation of the PoP of light is carried out by the twisted alignment of LC molecules where the electric vector of the polarized light rotates along the nematic director. A twisted nematic-LC (TNLC) cell works in a binary state, *i.e.*, either on (polarization rotation at twist angle) or off (no polarization rotation) state. The twist (rotation) angle is fixed for a given TNLC cell; therefore, it cannot be realized as a continuous polarization rotator [18]. A TNLC, satisfying the Mauguin's regime ($\alpha << 2\pi(\Delta nd)/\lambda = \Delta\phi$) can rotate PoP where the twist coefficient ($\alpha$) is much smaller than the phase retardation coefficient ($\Delta\phi$) [18,19]. To control the plane of polarized light continuously and precisely, we need to maintain the spatial alignment of the LC molecules [15]. Several attempts have been made in recent decades to realize continuous PR using LC. Ren *et al*. described a PR, where orthogonal rubbing treatment on a single substrate is established [20] in which they have shown a rotation of 0º to 90º. Further, a hybrid-aligned LC-based PR (LC-PR) [21] and a wavelength-independent tunable LC-PR are reported [22]. In addition, an optically controllable PR device is proposed using cholesteric LC by pitch modulation [23]. A PR using an electrically controlled LC waveguide is also documented [24]. In these devices, the continuous rotation of PoP from 0° to 90° is achieved. Recently, an electro-tunable achromatic PR using hybrid aligned LC with a rotation angle of 90º [25] (or 180º in tandem geometry) and an achromatic PR using dual-frequency LC [26] have been documented with rotation angles up to 180º. In these reports, the geometries of the proposed devices are complicated, and as a result, the fabrication process becomes very complicated. There is a significant requirement for a continuous PR with a broader tuning range of the PoP, which is easy to fabricate and applicable in wide optical applications.

An LC-based $\Theta$-cell converts linearly polarized light into radial and azimuthal polarization [27]. Here, we present a continuous polarization rotator using the $\Theta$-cell which is further demonstrated as a plane of polarization detector. A continuous rotation angle of ~160º is achieved that varies from 6.4º is 166.2º. For the detection of plane of polarization, a system is established using $\Theta$-cell, polarizer and camera that detects the PoP through imaging technique. The numerical simulation of linear polarization rotation is done using Mueller matrix formalism.

## 2. Design and Working Principle

Figure 1(a) shows the complete schematic of the proposed PR. The device comprises of two ITO-coated glass substrates of dimension $2 \times 2$ cm$^2$. The bottom substrate is treated with planar alignment, whereas circular grooves are created on the top substrate. The nematic 5CB LC is infiltrated between the substrates utilizing capillary action to realize LC $\Theta$-cell. It exhibits spatially varying twist angles across the LC cell, and the twist angle is always $\leq$ 90º. Like a TNLC- cell, this device rotates the plane of polarization through the principle of polarization guiding effect, where electric vector of the light follows the twist of LC. When the LC-PR is translated with respect to the optical beam, the light beam faces a continuously varying TN state [28]. Thus, the translation of the cell results in the continuous rotation of PoP. The twist angle of molecules at a particular position determines the angle of polarization rotation. The twisting of the molecules at different locations of the cell is shown schematically in Figure 1(b). Due to the opposite twist in the two halves of a circle, a geometric phase arises, and a disclination line appears along the rubbing direction. The twist is found to be maximum nearby the disclination line.

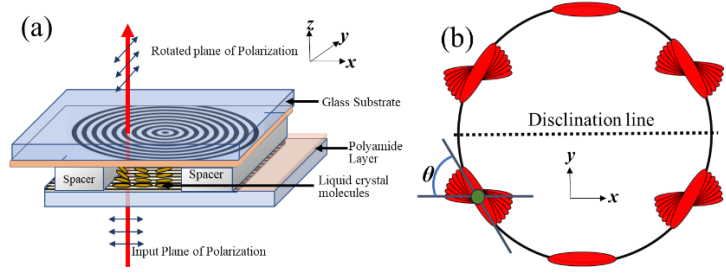

Figure 1. (a) Schematic geometry of the proposed liquid crystal-based based Θ-cell, (b) Top view of the twisted molecules at different locations of the cell in the *x-y* plane observed from the *z*-axis.

## 3. Device construction and analysis

First, we have examined the polarization rotation angle through the geometrical constructions of the TN state at different positions of the device. We designate the two substrates as input and output substrates where the light beam is incident on the input substrate and transmitted through the output substrate. The input substrate partakes planar alignment, whereas the output substrate is treated with circular alignment such that LC molecules form spatially varying twisted states. The laser beam propagates in the *z*-direction, and the device is mechanically translated in the *xy* plane to realize variable TN states. Initially, the device is translated in *x*-direction (*x*-scan) with respect to the focused laser beam.

The input laser beam is linearly polarized with its PoP parallel to the director of LC molecules (*x*-axis) on the input substrate. When the beam passes through the LC cell, the LC molecules on the output substrate must be aligned at some angle according to the circular alignment. The angle of the output molecules determines the twist angle at the beam propagation position. For instance, the position of beam propagation is indicated by a green point in Figure 1(b). The tangent at this point gives the direction of orientation of the LC molecule attached to the output substrate. The angle of orientation *θ* of LC molecules on the output substrate with respect to the input substrate can be calculated using the slope of the tangent. This angle (*θ*) is also referred to as the twist angle faced by the optical beam at the beam propagation position. Thus, the beam observes a continuously varying twist when we translate the device in either *x* or *y* direction, which results in a continuous rotation of the PoP of output beam. The geometrical design is analyzed by using the equation of a circle, $x^2 + y^2 = r^2$, where coordinates *x* and *y* are the coordinates of the point from where the laser is passing, and *r* is the radius of this circle. Thus, the total rotation in the PoP due to twist can be calculated by measuring the slope angles. By differentiating the equation of the circle, the angle of slope *θ* is given by equation (1):

$$\theta(x,y) = \tan^{-1}(x/y) \qquad (1)$$

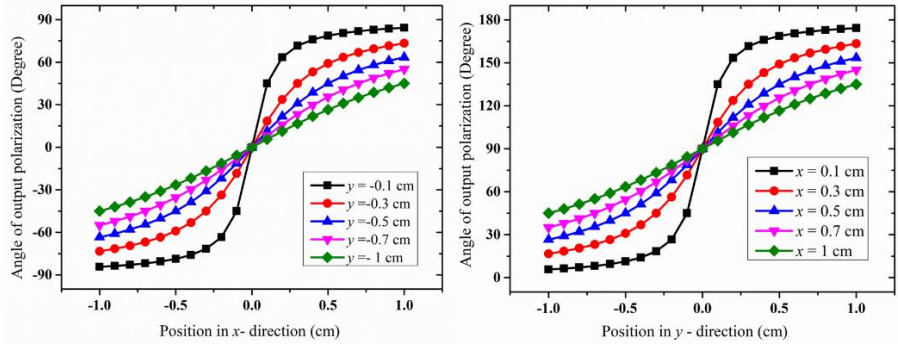

Figure 2. Observation of the plane of polarization in Θ-cell (a) Polarization distribution in the *x*-direction at different values of *y* (b) Polarization distribution in the *y*-direction at different values of *x*.

For the *x*-direction scan, *x* varies from -1 to +1 cm, keeping *y* constant. For different *y* values, the variation of angle ($\theta$) of PoP is illustrated in Figure 2(a). Figure 1 reveals that the geometry exhibits maximum twist at the extreme values of *x* and zero twisting effect at $x = 0$. At $y = -0.1$ cm, as the value of *x* increases from -1 cm to zero, the polarization rotation angle increases from -84.28° to 0°. At $x = 0$, the polarization rotation is also 0 because LC molecules lie in planar alignment. For the positive value of *x* from 0 to +1 cm, the polarization rotation angle increases up to +84.28°.

Similarly, for $y = -0.2$ cm, the polarization rotation angle increases as the *x* varies from -1 to +1 cm, but the maximum twist is smaller than the previous scan. It is also depicted in equation (1) and Figure 2(b) that the larger the distance of the scanning line from the center (*y*), the smaller will be the maximum twist at the extreme value of *x*. A similar trend is observed during the *y* scan, keeping *x* constant. We obtained the numerical results for the *y*-scan, as shown in Figure 2(b). As the value of *y* increases from -1 cm to 1cm, the polarization rotation angle is observed to be continuously varying from 5.8° to 174.3° at $x = 0.1$ cm. The polarization rotation angle of 90° is obtained due to the 90° twist faced by the incoming optical beam at $y = 0$. Analogous to the previous case, the range of polarization rotation angle decreases as the distance of the scanning line from the center increases. Thus, a position-dependent polarization distribution is seen in the device. Further, we constructed the Mueller matrix (MMx) for the proposed polarization rotator and then the mathematical analysis is performed. MMx is a 4×4 transformation matrix, which shows the effect of a polarizing component when an optical beam propagates through it. The standard MMx for a rotator [14] is given by equation (2):

$$M_R(\theta) = \begin{bmatrix} 1 & 0 & 0 & 0 \\ 0 & \cos 2\theta & \sin 2\theta & 0 \\ 0 & -\sin 2\theta & \cos 2\theta & 0 \\ 0 & 0 & 0 & 1 \end{bmatrix} \qquad (2)$$

Substituting the values of $\theta$ values from equation (1) into equation (2), the MMx for the presented LCPR can be written as:

$$M_{\{LCPR\}}(x,y) = \begin{bmatrix} 1 & 0 & 0 & 0 \\ 0 & \cos(2\tan^{-1}(-x/y)) & \sin(2\tan^{-1}(-x/y)) & 0 \\ 0 & -\sin(2\tan^{-1}(-x/y)) & \cos(2\tan^{-1}(-x/y)) & 0 \\ 0 & 0 & 0 & 1 \end{bmatrix} \quad (3)$$

The rotational effect of this matrix is observed in the rotation of incident linearly polarized light. Since the Stokes parameters are the blueprint for the determination of the PoP; therefore, we first calculate the output Stokes parameters. The output of Stokes parameters is calculated by multiplying equation (3) with the Stokes column vector of $x$-polarized light as given in equation (4) [13,14]:

$$S''(x,y) = \begin{bmatrix} 1 & 0 & 0 & 0 \\ 0 & \cos(2\tan^{-1}(-x/y)) & \sin(2\tan^{-1}(-x/y)) & 0 \\ 0 & -\sin(2\tan^{-1}(-x/y)) & \cos(2\tan^{-1}(-x/y)) & 0 \\ 0 & 0 & 0 & 1 \end{bmatrix} \begin{bmatrix} 1 \\ 1 \\ 0 \\ 0 \end{bmatrix} \quad (4)$$

The resulting output $S''(x, y)$, is the desired Stokes column vector, which reveals the Stokes parameters ($S_0$, $S_1$, $S_2$, and $S_3$) of the polarized optical beam. Utilizing the MATLAB Platform, the SoP of the light beam is plotted in terms of the electric field vector components using the calculated Stokes parameters. The simulated results extracted from equation (4) are shown in Figure 3.

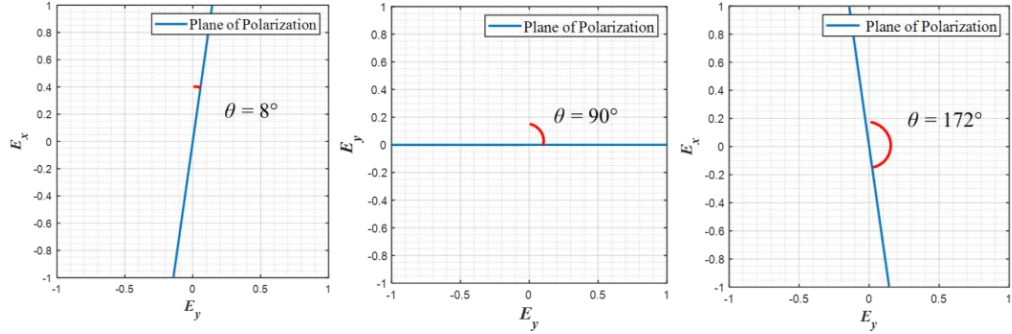

Figure 3. Simulated result of polarization state in Θ-cell using Mueller matrix formalism. Observations are taken with $x = 0.1$ cm (a) at $y = -0.7$ cm (b) at $y = 0$ cm (c) at $y = 0.7$ cm

The angle of PoP ($\theta$) of light is calculated at three cell positions at a scanning line of $x = 0.1$ cm. At $y = -0.7$ cm, the angle of PoP is observed as 8° (see Figure 3(a)), which increases as $y$ increases, and it becomes 90° at $y = 0$, as shown in Figure 3(b). On further cell movement, the PoP becomes 172° at $y = 0.7$ cm, as depicted in Figure 3(c). These simulated results are further substantiated using experimental observations.

## 4. Fabrication

The detailed fabrication procedure of the device is presented in Figure 4. First, we clean the ITO-coated glass substrates with acetone, isopropyl alcohol, and deionized water. The cleaned substrates were dehydrated using a heating plate at a temperature of 100°C for two minutes. The polyamide (PA) layer of nylon 6,6 is spin-coated on both the substrates using a spin coater at 4000 rpm for 30 seconds, as shown in Figure 4(a and b). This PA layer is baked for 20 minutes in the oven at 120°C. The first substrate is treated with straight grooves created by unidirectional mechanical rubbing using a velvet cloth, as shown in Figure 4(c). The circular grooves are created on the second substrate (see Figure 4(d) by contacting the velvet cloth to the spinning substrate at 5000 rpm. However, advanced methods such as the photoalignment technique result in a better alignment of LC [29]. In this work, mechanical rubbing is used only for the active demonstration of the device. Both the substrates are assembled using UV-curable adhesive by keeping the spacers of thickness 6 µm. For the application of an electric field, electrical connections are made on the ITO surface using indium ingots (see Figure 4(e)). The 5CB LC is filled in the assembled device utilizing capillary action with the help of a medical syringe needle. The LC molecules get aligned straight on the bottom substrate and circular on the top substrate. The LC alignment is confirmed using a polarizing optical microscope (POM) by recording the transmission pattern under crossed polarizers.

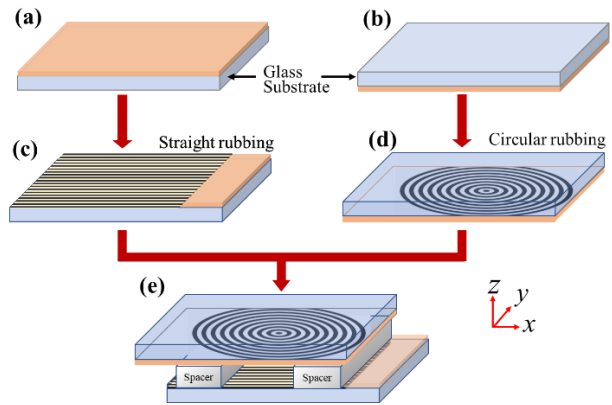

Figure 4. Fabrication steps involved in the development of LC Θ-cell: (a, b) ITO-coated glass substrate with a polyamide coating, (c) linear rubbing, (d) circular rubbing, (e) assembled LC-cell with the help of the Mylar spacers.

## 5. Experimental Characterization Setup

The experimental setup to characterize PR is portrayed in Figure 5. The setup consists of a linearly polarized He-Ne laser as the light source, which emits a collimated beam of λ = 633 nm that propagates in the $z$-direction. A pair of microscope objectives (20X) $MO_1$ and $MO_2$ are used to focus and collimate the laser beam. The LCPR is placed at the focal point of $MO_1$. The collimated beam after the $MO_2$ propagates through a quarter wave plate QWP ($\lambda/4$) and a polarizer (Thorlabs). The combination of a polarizer and QWP is used as the Stokes polarimeter [13], which determines the polarization state of light. Here, we used the classical method to measure Stokes parameters. We measured the optical power using a photodetector further connected to the power meter. A function generator is used to apply the electric field across the LCPR cell. The intensity after passing ($\lambda/4$) and polarizer $P_1$ is given by equation (5) [13, 14].

$$I(\alpha, \phi) = 0.5(S_0 + S_1 \cos 2\alpha + S_2 \sin 2\alpha \cos\phi + S_3 \sin 2\alpha \sin\phi) \quad (5)$$

where $\alpha$ and $\phi$ are the angles made by the transmission axis of the polarizer and the fast axis of QWP with the incident beam polarization vector, respectively. The Stokes parameters are calculated using equation (5) by measuring the optical field intensities [13]. The first three Stokes parameters $S_0$, $S_1$, and $S_2$, are estimated using polarizer $P_1$ with its transmission axis at 0°, 90°, and 45°, respectively, without introducing the QWP in the optical setup. The fourth Stokes parameter, $S_3$, is measured by introducing a QWP plate with a fast axis at 0° placed before $P_1$ with pass axis at 45°.

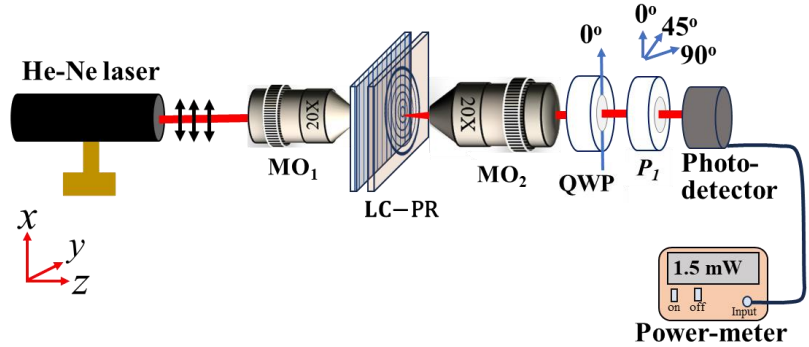

Figure 5. Schematic of the experimental setup for detection of the PoP.

## 6. Results and Discussion

*6.1 Characterization of Polarization Rotator*

To characterize the polarization rotator, a linearly polarized laser beam is launched on the device such that the electric vector of the polarization remains parallel to the straight-aligned LC molecules on the rear substrate (towards the source). As explained earlier, the LC molecules on the back substrate are circularly aligned, due to which the device exhibits continuously varying TN-states. The polarization of the incident beam gets rotated in accordance with the twist faced by the optical beam. Figure 6(a) illustrates the schematic of the device to be scanned by an *x*-polarized laser beam in the *y*-direction at a fixed value of *x* = 0.1 cm.

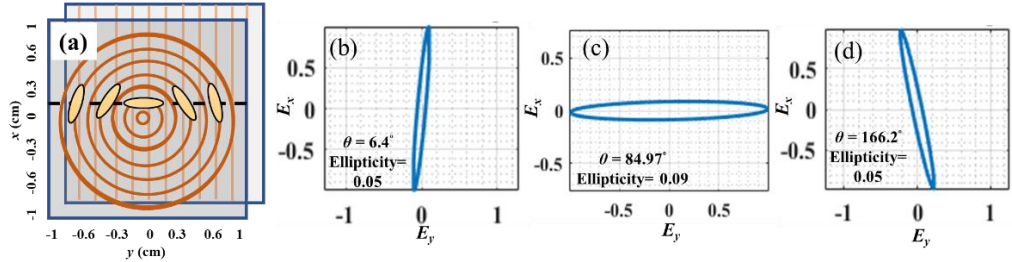

Figure 6. (a) Schematic top view of the device showing a scanning line twisting of top substrate molecules, (b) SoP of the transmitted light at *y* =-0.7 cm, (c) 0, and (d) + 0.7 cm.

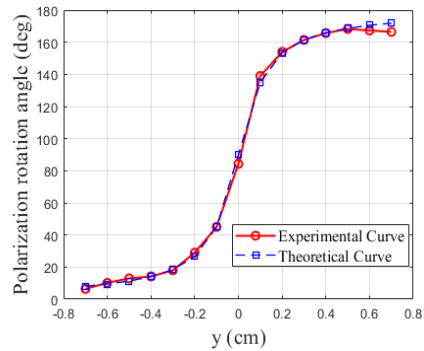

Figure 7. Variation of polarization rotation angle on the scanning in *y*-direction and comparison with theoretical results.

cm. The polarization gets rotated in accordance with the orientation of the molecule on the top substrate. The MATLAB platform is used to examine the polarization state of light by using the Stokes polarimetry method utilizing equation (5). The center of the circles is considered as the reference for measuements, as shown in Figure 6(a). We obtained the output polarization state from the Stokes polarimetry at three different positions corresponding to *y* =-0.7, 0, and +0.7, as shown in Figure 6(b), 6(c), and 6(d), respectively, at *x* = 0.1 cm. Here, the orientation angle (OA), *i.e.*, the angle of polarization rotation, is obtained as 6.4°, 84.9° and 166.2°. It is observed that the output light beam retains some degree of ellipticity instead of perfect linear polarization, which is due to the circularly aligned LC molecules at the output substrate. The light beam faces a spatially varying twist in the illuminated area. Therefore, the output light carries some degree of ellipticity, which can be further reduced by minimizing the laser beam spot passing through the beam. In order to observe continuous rotation in the output polarization, the LC cell is translated in *y*-direction, and the output polarization is recorded at an interval of 0.1 cm. The device scanning starts from $y = -0.7$ cm to $y = 0.7$ cm. It is observed that the plane of polarization also rotates continuously as we move the cell. The continuous rotation of the PoP of the optical beam is illustrated in Figure 7, along with the theoretical values.

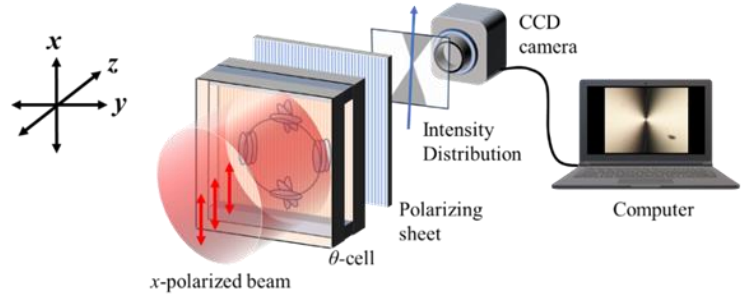

Figure 8. Experimental setup for the detection of the angle of the plane of polarization.

Theoretical and experimental observations are found to be in good agreement. Thus, the PoP of the optical beam rotated continuously from 6.4° to 166.2° at a fixed value of $x = 0.1$ cm. Furthermore, we implement this device as a plane of polarization detector described in the following section.

*6.2 Plane of polarization detector*

The fabricated LC Θ-cell is used to detect the angle of the plane of a linearly polarized light. The devices operate based on the principle that light passes through a polarizer when its electric field vector is parallel to its pass axis and blocked when perpendicular. Often, the plane of polarization is detected using linear polarizers employing Malus' law and Stoke's polarimeters [13]. The linear polarizer includes the mechanical movement of the components, which introduces backlash errors in the analysis and shorter the lifetime of the device. The Stoke's polarimeter requires multiple optical components and several sets of precise power measurements. These limitations make the measurement of Stoke's parameters an expensive, cumbersome, and sensitive process for polarization detection. Here, we placed the presented Θ-cell in reverse mode to detect the plane of a linearly polarized light. The experimental setup for the plane of polarization detection is depicted in Figure 8. The linearly polarized light enters the LC cell through the substrate with circular alignment and exits the straight-aligned substrate. As explained in the previous section, the propagated light through the LC cell acquires polarization modulation with different twist angles. The plane of the polarization gets rotated by 90° at the TN position and remains the same at the planar positions. A linear polarizer is used at the cell output to block the rotated components. The polarizer transmits the light wave components, which have electric vector in the direction of its optic axis. Thus, a dark lobe is observed in the intensity pattern, which corresponds to the blocked components by the polarizer. The axis of this dark-intensity lobe is parallel to the direction of the incident plane of polarization.

The device uses a CCD camera that captures the transmitted light passing through the devices and displays the intensity distribution. The obtained dark intensity distribution forms a Dumbbell-shaped pattern, as shown in Figure 8. The circular line intensity profile of the intensity distribution is plotted using the azimuthal tool in Origin9 software. The intensity patterns acquired for the different incident PoPs are shown in Figure 9(a-j). It can be seen that the dark intensity lobe rotates anti-clockwise with the rotation of the angle of PoP of incident light (see supplementary video). A circular (azimuthal) line is drawn on the obtained patterns, and intensity variation is investigated as a function of the azimuthal angle which is swept from 0° to 360°. The intensity variation as a function of the azimuthal angle is shown in Figure 10(a). When the input plane of polarization is at 0°, the black curve shows a dip at 0°, corresponding to the minimum intensity. As the angle of the PoP increases, the dip of the intensity curve shifts towards the higher angles. The minima of each curve are extracted and plotted as a function of the angle of incident PoP, as shown in Figure 10(b). The detected angle of PoP accurately predicts the incident PoP with high precision using a single shot of imaging technique.

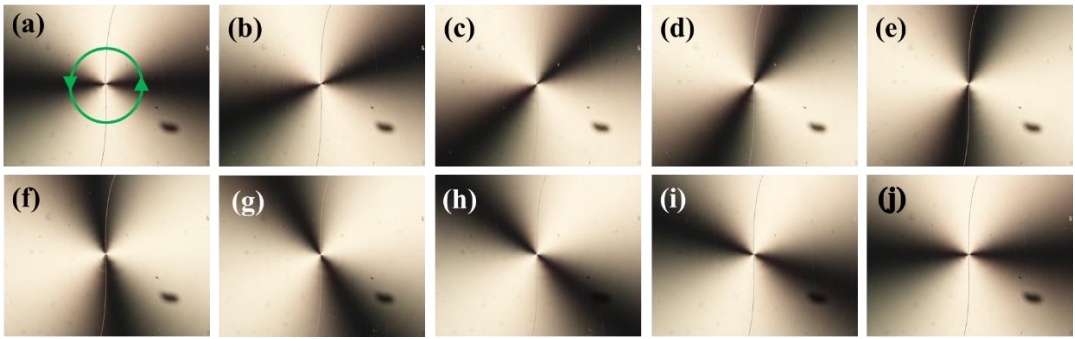

Figure 9. Intensity distribution of the transmitted light through Θ-cell and polarizer for an incident polarization angle of (a) 0°, (b) 20°, (c) 40°, (d) 60°, (e) 80°, (f) 100°, (g) 120°, (h) 140°, (i) 160°, (j) 180°.

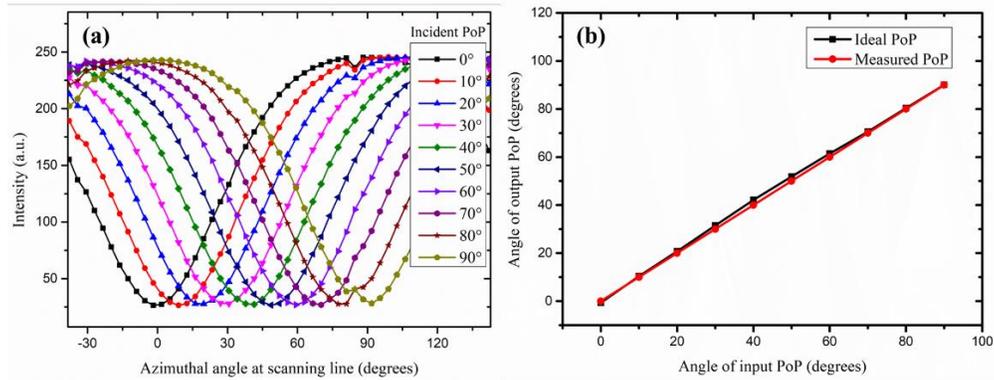

Figure 10. (a) Variation of intensity measured along a circular line drawn on the output intensity distribution, (b) Variation of output angle of polarization with input angle of polarization.

## 7. Conclusion

We have demonstrated a continuous rotation of PoP in LC-based Θ-cell configuration. The device consists of a circular and straight alignment of LC molecules anchored on two ITO-coated glass substrates and is characterized by probing a He-Ne laser (633 nm). The presented PR is mathematically constructed and analyzed using the Mueller matrix. The PoP is detected experimentally by using the Stokes polarimetry technique. The PoP of the probe beam undergoes continuous rotation from 6.4º to 166.2º during a straight-line scan of the device. The experimentally observed results agree with the calculated values. In the reverse mode, the presented device is demonstrated as an accurate plane of polarization detector through the imaging of transmitted intensity distribution. The experimentally detected plane of polarization is observed to be the same as the incident plane of polarization. The development of the device includes a simple spin-coating facility with cost-effective materials. The proposed device presents a smart method of polarization detection, which can have potential applications in polarization-based experiments, medical imaging, and advanced optical instrumentations.

**Funding.** Government of India, Ministry of Defence, Defence Research and Development Organization, New Delhi (DFTM/03/3203/P/01/JATCP2QP-01).

**Acknowledgments.** The authors thank the Indian Institute of Technology Delhi, New Delhi, and the Defence Research & Development Organization, India, for project funding and research facilities.

**Disclosures.** The authors declare no conflicts of interest.

**Data availability.** Data underlying the results presented in this paper are not publicly available at this time but may be obtained from the authors upon reasonable request.

## References

1. Jofre M, Anzolin G, Steinlechner F, Oliverio N, Torres J, Pruneri V and Mitchell M 2012 Fast beam steering with full polarization control using a galvanometric optical scanner and polarization controller *Opt. Express* **20** 12247.
2. Pishnyak O, Kreminska L, Lavrentovich O D, Pouch J J, Miranda F A and Winker B K 2005 Liquid Crystal Digital Beam Steering Device Based on Decoupled Birefringent Deflector and Polarization Rotator *Mol. Cryst. Liq. Cryst.* **433** 279.
3. Demos S and Alfano R 1997 Optical polarization imaging *Appl. Opt.* **36** 150.
4. Solomon J E 1981 Polarization imaging *Appl. Opt.* **20** 1537.
5. Panchal R and Sinha A 2023 Electrically controlled continuous laser beam steering in a liquid crystal based electro-optic waveguide *Opt. Laser Technol.* **158** 108816.
6. Ram B B, Senthilkumaran P and Sharma A 2017 Polarization-based spatial filtering for directional and nondirectional edge enhancement using an s-waveplate *Appl. Opt.* **56** 3171.
7. Lesoine J F, Lee J Y, Krogmeier J R, Kang H, Clarke M L, Chang R, Sackett D L, Nossal R and Hwang J 2012 Quantitative scheme for full-field polarization rotating fluorescence microscopy using a liquid crystal variable retarder *Rev. Sci. Instrum.* **83** 053705.
8. Pal S K, Ruchi and Senthilkumaran P 2017 Polarization singularity index sign inversion by a half-wave plate *Appl. Opt.* **56** 6181.
9. Darsht M Y, Goltser I, Kundikova N and Zel'Dovich B Y 1995 Adjustable half-wave plate *Appl. Opt.* **34** 3658.
10. Simon R and Mukunda N 1989 Universal SU(2) gadget for polarization optics *Phys. Lett.* A **138** 474.
11. Ghatak A 2009 *Optics* (Avenue of the Americas: McGraw-Hill).
12. Hecht E 2017 *Optics*, (Harlow: Pearson Education Limited).
13. Goldstein D H 2017 *Polarized light* (Boca Raton: CRC press).
14. Al-Mahmoud M, Hristova H, Coda V, Rangelov A A, Vitanov N V and Montemezzani G 2021 Non-reciprocal wave retarder based on optical rotators combination *OSA Contin.* **4** 2695.
15. Moreno I, Davis J, Hernandez T, Cottrell D and Sand D 2012 Complete polarization control of light from a liquid crystal spatial light modulator *Opt. Express* **20** 364.
16. Peinado A, Lizana A and Campos J 2014 Use of ferroelectric liquid crystal panels to control state and degree of polarization in light beams *Opt. Lett.* **39** 659.
17. Gayaprasad and Kanseri B 2023 Degree and state of polarization control using brewster's law in a nematic liquid crystal *Opt. Laser Technol.* **157** 108705.
18. Saleh B E and Teich M C 2019 *Fundamentals of Photonics* (Hoboken: John Wiley & Sons)
19. Chung T-Y, Tsai M-C, C-K, Li J-H and Cheng K-T 2018 Achromatic linear polarization rotators by tandem twisted nematic liquid crystal cells *Sci. Rep.* **8** 13691.
20. Ren H and Wu S-T 2007 Liquid-crystal-based linear polarization rotator *Appl. Phys. Lett.* **90** 121123.

21. Yang F, Ruan L, Jewell S and Sambles J 2007 Polarization rotator using a hybrid aligned nematic liquid crystal cell *Opt. Express* **15** 4192.
22. Aharon O and Abdulhalim II I S 2010 Liquid crystal wavelength-independent continuous polarization rotator *Opt. Eng*. **49** 034002.
23. Liu C-K, Chiu C-Y, Morris S M, Tsai M-C, Chen C-C and Cheng K-T 2017 Optically controllable linear-polarization rotator using chiral-azobenzene-doped liquid crystals *Materials* **10** 1299.
24. Li T, Chen Q and Zhang X 2018 Electrically controlled polarization rotator using nematic liquid crystal *Opt. Express* **26** 32317.
25. Guo D-Y, Chang L-M, Chen C-W, Li C-C, Jau H-C, Wang C-T, Kuo W S and Lin T-H 2021 Electrotunable achromatic polarization rotator *Optica* **8** 364.
26. Chang L M, Feng T-M, Lin K-W, Tseng H-Y, Li C-C, Guo D-Y, Jau H-C, Wang C-T and Lin T-H 2022 Electrotunable 180° achromatic linear polarization rotator based on a dual-frequency liquid crystal *Opt. Express* **30** 4886.
27. Stalder M and M Schadt 1996 Linearly polarized light with axial symmetry generated by liquid-crystal polarization converters *Opt. Lett.* **21** 1948.
28. Vasnetsov M, Pas'ko V and Kasyanyuk D 2011 Observation of polarization conflict caused by geometrical phase in a twisted nematic liquid crystal cell *Opt. Lett.* **36** 2134.
29. Liu C-K, Liao S-H, Haung C-T and Cheng K-T 2023 Fabrication of azimuthally/radially symmetric liquid crystal plates using two-step photoalignments *Opt. Express* **31** 21962.